\documentclass[12pt]{article}
\usepackage{amsfonts,amssymb}

\renewcommand{\baselinestretch}{1.4}
\newcommand{\R}[1]{\mathbb{R}^{#1}}

\def\be{\begin{equation}}
\def\ee{\end{equation}}
\def\bear{\begin{eqnarray}}
\def\eear{\end{eqnarray}}

\def\bS{\mathbf{S}}

\def\Vol{\mathrm{Vol}}
\def\vol{\mathrm{vol}}

\def\bi{\bibitem}

\begin{document}

\begin{titlepage}

\begin{flushright}
hep-th/0210126\\
NSF-ITP-02-155
\end{flushright}
\vfil

\begin{center}
{\huge The Hydrodynamics of M-Theory}\\

\end{center}

\vfil
\begin{center}
{\large Christopher P. Herzog}\\
\vspace{1mm}
Kavli Institute for Theoretical Physics,\\
University of California, Santa Barbara, CA  93106, USA\\
{\tt herzog@kitp.ucsb.edu}\\
\vspace{3mm}
\end{center}

\vfil

\begin{center}

{\large Abstract}
\end{center}

\noindent
We consider the low energy limit of a stack of $N$ M-branes at finite 
temperature.  In this limit, the M-branes are well described, 
via the AdS/CFT correspondence, in terms of classical solutions
to the eleven dimensional supergravity equations of motion.
We calculate Minkowski space two-point functions on these M-branes
in the long-distance, low-frequency limit, i.e. the hydrodynamic
limit, 
using the prescription of Son and Starinets
[{\tt hep-th/0205051}].  
From these Green's functions for the R-currents and for
components of the stress-energy tensor, we extract two kinds
of diffusion constant and a viscosity.  The $N$ dependence
of these physical quantities 
may help lead to a better understanding of M-branes.

\vfil
\begin{flushleft}
October 2002
\end{flushleft}
\vfil
\end{titlepage}
\newpage
\renewcommand{\baselinestretch}{1.1}  

\renewcommand{\arraystretch}{1.5}

\section{Introduction}

The interacting, superconformal field theories (SCFT) living on
a stack of $N$ M2- or M5-branes are not well understood.  
An improved understanding of these M-branes should 
lead eventually to a better
understanding of M-theory itself, a theory that encompasses
all the different super string theories and is one of the best
hopes for a quantum theory of gravity.  While the full M-brane
theories remain mysterious, the low energy, large $N$ behavior is 
conjectured to be described well, via the AdS/CFT correspondence
\cite{jthroat, EW, GKP},
by certain classical solutions to eleven dimensional
supergravity equations of motion.  Recent work on
AdS/CFT correspondence 
by Son, Starinets, and Policastro \cite{SS, PSS} provides a 
prescription for writing Minkowski space two-point functions
for these types of theories.  We take advantage of this
prescription to calculate 
viscosities and diffusion constants for M-brane theories in this
low energy limit, thus generalizing the work of \cite{PSS} 
for D3-branes.

Policastro, Son, and Starinets \cite{PSS} used their
Minkowski space prescription to investigate the
low-frequency, long-distance, finite temperature regime
of the D3-brane theory.
There is lore
that the long-distance, low-frequency behavior of any interacting
theory at finite temperature can be described well by fluid 
mechanics (hydrodynamics) \cite{Landau}.  Although not
rigorously proven, the idea is well supported by physical
intuition about macroscopic systems.  Hydrodynamics in turn
provides rigorous constraints for the form of Minkowski
space correlation functions.  Once a few
viscosities, diffusion constants, and other transport coefficients
are known, the two-point functions are completely
fixed \cite{KadMart}.

Indeed, the authors of \cite{PSS} found that the form
of the Green's functions calculated from
supergravity was completely consistent with
hydrodynamics for this D3-brane theory.  Moreover,
from these Green's functions, they were able to extract
a transport coefficient (a shear viscosity) and diffusion constants.  
The authors' prescription for Minkowski space Green's functions
is a modification
of the prescription for calculating Euclidean space 
Green's function developed in \cite{EW, GKP}. 
Both prescriptions use the gravitational description of
the low energy theory on the brane for calculating
correlation functions.

It is true that these same Green's functions
on a D3-brane can be calculated in a more traditional way using 
weakly coupled gauge theory \cite{Sonfriends}.  
The low energy theory living on a stack of $N$ D3-branes is described
via the AdS/CFT correspondence, alternately as ${\mathcal N}=4$
$SU(N)$ super Yang Mills or as type IIB supergravity in a 
$AdS_5 \times \bS^5$ background. 
While the gravity
calculation is good at strong `t Hooft coupling $\lambda$
(large curvature),
the gauge theory calculation is good only for small
$\lambda$.  However, the $N$ and temperature dependence of these
transport coefficients and diffusion constants 
should be and indeed is
universal and 
independent of the method of calculation.  Remarkably,
the gravity calculation appears to be technically simpler
than the equivalent calculation for weakly coupled gauge theory.

In contrast, the M-brane theories have no such alternate
gauge theory description.  However, there are still
low-energy supergravity solutions from which we can
extract analogous Minkowski space Green's functions.  
Moreover, we can hope that M-branes, like D3-branes,
have a hydrodynamic regime. 

Indeed, based on the Minkowski space prescription of Son and Starinets
\cite{SS}, in the low-frequency, long-distance,
finite temperature regime, we find that these M-brane theories have
two-point functions which are completely consistent with
a hydrodynamic interpretation.  Generalizing
\cite{PSS}, we calculate two-point functions for the
conserved R-symmetry current and for components of the
stress-energy tensor for these M-branes.
Moreover, from
these Green's functions we are able to extract 
corresponding diffusion constants and also the viscosity.
The $N$ dependence of these physical quantities
may lead to a better understanding of 
M-brane theories.

It should be emphasized that the Son and Starinets
prescription \cite{SS} for calculating Minkowski space correlators
is not completely justified, and that this paper provides
some limited additional evidence for their prescription.  
Turning the logic of the last two paragraphs around, we can argue
that on general grounds we expected the M-branes to have
a hydrodynamic description.  Thus, it is reassuring
that the prescription of \cite{SS} does indeed 
produce Green's functions with the appropriate behavior.
Also, we get results which are internally consistent.   
On general hydrodynamic grounds, we expect that
$D = \eta/ (\epsilon + P)$ where $D$ is the diffusion constant
calculated from the stress energy tensor,
$\eta$ the viscosity, $\epsilon$ the energy density, 
and $P$ the pressure.
We check that this equation holds both for the M2-branes
and M5-branes.

We begin by reviewing some essential facts about the 
non-extremal M2- and M5-brane backgrounds.  As we are
working at finite temperature, the extremal
$AdS_4 \times \bS^7$ and $AdS_7 \times \bS^4$ 
supergravity solutions are not adequate.
We need their nonextremal generalizations where we 
can associate a Hawking temperature to the horizon.

Next, we consider R-current correlation functions.  Both
M2- and M5-brane supergravity solutions have an R-symmetry
which one can think of roughly as the rotational symmetry
of the transverse sphere.  From the form of the thermal
R-current two-point functions, we extract a corresponding
R-charge diffusion constant.

Having warmed up with the R-current correlators, we proceed
to the more complicated example of stress-energy tensor
two-point functions.  From components of these two-point
functions, we extract a diffusion constant and a viscosity for
each M-brane theory.

Finally, we end with some comments about the $N$ dependence
or lack thereof of the various diffusion constants
and transport coefficients calculated.  
The motivation for this work came in large part out of 
the hope that some of these $N$ dependences might shed
light on the underlying M-brane theories.

\section{The Nonextremal Eleven Dimensional Supergravity Backgrounds}

We begin by reviewing some essential facts about the non-extremal
M2- and M5-brane supergravity solutions.  Roughly speaking, these
solutions represent what happens when a stack of M-branes is 
placed in flat 11-dimensional space and given some finite temperature
$T$.  The 11-dimensional space, close to the M-branes, separates
into a product of a sphere and an asymptotically anti-de Sitter space.
A horizon with Hawking temperature $T$ forms.
These supergravity solutions
are solutions to the equations of motion following from the 
eleven dimensional supergravity action 
\cite{Cremmer} 
\begin{equation}
\frac{1}{2\kappa_{11}^2} \int d^{11} x (-g)^{1/2} {\mathcal R}
-
\frac{1}{4\kappa_{11}^2} \int \biggl( F_4 \wedge {\star} F_4 +
\frac{1}{3} A_3 \wedge F_4 \wedge F_4 \biggr).
\label{11dsugra}
\end{equation}
where $\kappa_{11}$ is the gravitational coupling strength, and 
$dA_3=F_4$.

%

\subsection{M5-brane}

For the M5-brane, the nonextremal metric is
\be
ds^2 = H(r)^{-1/3} \left[ -f(r) dt^2 + d \vec{x}^2 \right]
+ H(r)^{2/3} \left[ \frac{dr^2}{f(r)} + r^2 d\Omega_4^2 \right]
\ee
where $H(r) = 1 + R^3/r^3$ and $f(r) = 1 - r_0^3/r^3$.  The
quantity $d \vec{x}^2$ is a metric on flat, Euclidean $\R{5}$.
The term $d\Omega_4^2$ is the metric on a unit four sphere $\bS^4$.
The four form flux $F_4$ from the M5-branes threads this
$\bS^4$:
\be
F_4 = 3 R^3 \vol(\bS^4).
\ee  
The quantization condition on the flux implies that 
$N^3 \kappa_{11}^2 = 2^7 \pi^5 R^9$ where $N$ is the number
of M5-branes and $\kappa_{11}$ is the eleven dimensional
gravitational coupling strength \cite{Kleb}.

Taking the near-horizon limit $r \ll R$, according to the 
AdS/CFT prescription \cite{jthroat,EW,GKP} 
we can ``zoom in'' on the
five-brane theory dynamics:
\be
ds^2 \rightarrow \frac{r_0}{uR} \left[ -f(u) dt^2 + d\vec x^2 \right]
+\frac{R^2}{f(u)} \frac{du^2}{u^2} + R^2 d\Omega_4^2  \ .
\label{met7}
\ee
We have made the coordinate transformation $u = r_0 / r$.  Spatial infinity,
which is also now the boundary of an asymptotically 
anti-de Sitter space,
corresponds to $u=0$.  There is a horizon at $u=1$.
The Hawking temperature of this horizon is
\be
T = \frac{3}{4\pi} \frac{r_0^{1/2}}{R^{3/2}} \ .
\ee

Another important quantity characterizing this supergravity solution
is the entropy density, which one finds by multiplying the horizon area by
$2\pi / \kappa_{11}^2$ and dividing out by the volume of the gauge theory 
directions, $x^1, x^2, \ldots, x^5$ \cite{KlebTseyt}:
\be
S = \frac{2^7 \pi^3}{3^6}N^3 T^5 \ .
\label{S5}
\ee

We will use index conventions where $\mu, \nu, \ldots$ refer to
the asymptotically $AdS$ directions, $\alpha, \beta, \ldots$
index the M-brane directions only, and 
$i, j, \ldots$ index the spatial M-brane directions.

\subsection{M2-brane}

An analogous nonextremal supergravity solution exists for a stack of 
M2-branes in eleven dimensional space.  Now the metric
takes the form
\be
ds^2 = H(r)^{-2/3} \left[ -f(r) dt^2 + dx^2 + dy^2 \right] +
H(r)^{1/3} \left[ \frac{dr^2}{f(r)} + r^2 d\Omega_7^2 \right] \ 
\ee
where $H(r) = 1+R^6/r^6$ and $f(r) = 1-r_0^6/r^6$.
The four form field strength is easier to think of in a dual language
as a seven form field strength:
\be
{\star}F_4 = F_7 = 6 R^6 \vol(\bS^7) \ .
\ee
The quantization condition \cite{Kleb} on the field strength reveals that
$R^9 \pi^5 = N^{3/2} \kappa_{11}^2 \sqrt{2}$.

We can again take a near horizon limit, $r \ll R$, to find
\be
ds^2 \rightarrow \frac{r_0^4}{u^2 R^4} \left[
-f(u) dt^2 + dx^2 + dy^2 \right] + \frac{R^2}{4f(u)} \frac{du^2}{u^2} + 
R^2 d\Omega_7^2 \ ,
\label{met4}
\ee
where $u = r_0^2/r^2$.
The Hawking temperature is
\be
T = \frac{3}{2\pi} \frac{r_0^2}{R^3} \ ,
\ee
and the entropy density is \cite{KlebTseyt}
\be
S = \frac{8\sqrt{2} \pi^2}{27} N^{3/2} T^2 \ .
\label{S2}
\ee

\section{R-charge Diffusion for M2- and M5-branes}

The R-charge interactions are mediated in the bulk 
by a gauge field $F_{\mu \nu}^a$.
The starting point for calculating two-point R-charge correlators is 
the usual Maxwell action in 
the nonextremal backgrounds given above:
\be
S = -\frac{1}{4 g_{SG}^2}\int d^d x \sqrt{-g} F_{\mu\nu}^a {F^{\mu\nu}}^{\, a} 
\ .
\ee
Here $d$ is equal to 4 for the M2-branes and 7 for the M5-branes.  
The tensor $g_{\mu\nu}$ is a metric on the asymptotically 
anti-de Sitter space. 
The calculation is similar to the Euclidean calculations 
in \cite{Rcurrent}; however we work with a Lorentzian signature
and in a nonextremal background.
The constant $g_{SG}$ can be set by compactifying the eleven dimensional
supergravity action $(\ref{11dsugra})$ on a $\bS^4$ or $\bS^7$.  
We will ignore
this overall normalization for the Green's functions for now.
The reason is that 
our main interest in this section is the diffusion coefficient for
the R-current which can be obtained simply from the location of the pole
in the corresponding retarded Green's functions.
The overall normalization of the Green's function is also related to
the diffusion constant, but in a more complicated way via a Kubo-type
formula.

Our calculations closely follow \cite{PSS}.  We work in the gauge
$A_u=0$.  We use a Fourier decomposition
\be
A_\mu = \int \frac{d^{d-1} q}{(2\pi)^{d-1}} e^{-i\omega t + i {\bf
q}\cdot{\bf x}}
A_\mu(q,u) .
\ee
Rotational invariance in the spatial directions allows one to simplify
things further by choosing $q^0 = \omega$, $q^1=q$, and all other $q^i=0$.
The equations of motion for the $A_\mu$ are
\be
\frac{1}{\sqrt{-g}} \partial_\nu \left[
\sqrt{-g} g^{\mu \rho} g^{\nu \sigma} \left(
\partial_\rho A_\sigma - \partial_\sigma A_\rho \right) \right] = 0 \ .
\label{Aeom}
\ee
At this point, it becomes convenient to analyze the M2- and M5-brane
cases separately.

\subsection{R-charge and M5-branes}

The equations of motion (\ref{Aeom}) for $A_\mu$ reduce to
\bear
\omega_5 A_t' + f q_5 A_x' &=& 0 
\ , \label{A1} \\
A_t'' - \frac{1}{u} A_t' - \frac{1}{uf} \left(
\omega_5 q_5 A_x + q_5^2 A_t \right) &=& 0 
\ , \label{A2} \\
A_x'' - \frac{1}{u} A_x' + \frac{f'}{f} A_x' 
+\frac{1}{uf^2} \left( q_5 \omega_5 A_t + \omega_5^2 A_x \right) &=& 0
\ , \label{A3} \\
A_\alpha'' - \frac{1}{u} A_\alpha' + \frac{f'}{f} A_\alpha' + 
\frac{1}{uf^2} \left( \omega_5^2 - f q_5^2 \right) A_\alpha &=& 0
\label{A4} \ ,
\eear
where $t$ is $x^0$, $x$ is $x^1$, and $\alpha$ stands for any of the other 
$x^i$.  
We have introduced 
\be
\omega_5 \equiv  \frac{3}{4\pi T} \omega \; \; \; ;
\; q_5 \equiv \frac{3}{4\pi T} q \ .
\ee
The prime denotes the derivative with respect to $u$.
The peculiar combinations of
$q_5^2$, $\omega_5 q_5$, and $q_5^2$ in the equations guarantee gauge
invariance under the residual transformation 
$A_t \rightarrow A_t - \omega \Lambda$ and $A_x \rightarrow A_x + q \Lambda$.
The first three equations are dependent; equations (\ref{A1}) and
(\ref{A2}) imply equation (\ref{A3}).  

Because
of our choice of $q$-vector, 
the R-charge diffusion appears only in the $A_t$ and 
$A_x$ sector, and we 
begin with the first two equations, (\ref{A1}) and (\ref{A2}). 
These two equations can be combined to yield a single
equation for $A_t'$:
\be
A_t''' + \frac{f'}{f} A_t'' + \frac{\omega_5^2-q_5^2 f - ff'}{uf^2} A_t' = 0 
\ .
\ee
This second order equation for $A_t'$ 
does not appear to be analytically tractable.
However, a solution can be obtained perturbatively for small $q_5$ and
$\omega_5$.  Following \cite{PSS}, we determine the behavior of $A_t'$
near the singular point $u=1$.  Substituting 
$A_t' = (1-u)^\alpha F(u)$, where $F(u)$ is a regular function,
one finds that $\alpha^2 = -\omega_5^2/9$.  The ``incoming wave''
boundary condition described in \cite{SS} forces us to choose
$\alpha = - i \omega_5/3$.

Next, we solve for $F(u)$ perturbatively:
\be
F(u) = F_0(u) + \omega_5 F_1(u) + q_5^2 G_1(u) + 
{\mathcal O}(\omega_5^2, \omega_5 q_5^2, q_5^4 ) \ ,
\ee
where $F_0(u) = u C$ and
\bear
F_1(u) &=& i C \left[ (u-1) + \frac{u}{6} f_1(u)
-\frac{u}{\sqrt{3}} f_2(u)
\right] \ , \\
G_1(u) &=& C \left[ \frac{1}{2}(1-u) 
+\frac{u}{\sqrt{3}} f_2(u)
\right] \ ,
\eear
where
\be
f_1(u) = \ln \frac{1+u+u^2}{3} \; \; \; ; \;
f_2(u) = \tan^{-1} \left( \frac{1+2u}{\sqrt{3}} \right) - \frac{\pi}{3}
\ .
\ee
One of the two integration constants for $F_0(u)$ is set by requiring
that $F_0(u)$ is well-behaved at $u=1$.  The integration constants
for all higher $F_i(u)$ and $G_i(u)$ are set by requiring that these
functions vanish at $u=1$ as well.

The constant $C$ can be related to the boundary values $A_t^0$ and
$A_x^0$ using (\ref{A2}):
\be
C =\frac{\omega_5 q_5 A_x^0 + q_5^2 A_t^0}{i \omega_5 - \frac{1}{2} q_5^2} \ .
\ee
The pole in $C$ is the same pole that appears in the 
retarded Green's functions, as we will presently see.
Having obtained $A_t'$, $A_x'$ follows from (\ref{A1}).

The solution to (\ref{A4}) can be obtained in a similar fashion:
\be
A_\alpha = \frac{A_\alpha^0 (1-u)^{-i\omega_5/3} h(u)}{h(0)} 
+ 
{\mathcal O} (\omega_5^2, \omega_5 q_5^2, q_5^4)
\ee
where
\be
h(u) = 1+ i\omega_5 \left( 
\frac{1}{6} f_1(u) - \frac{1}{\sqrt{3}} f_2(u)
\right)  - q_5^2 \frac{2}{\sqrt{3}} f_2(u) \ .
\ee

The Green's functions can now be calculated from the terms in
the action which contain two derivatives with respect to $u$:
\bear
S &=& -\frac{1}{2g_{SG}^2} \int du \, d^6x \, \sqrt{-g} g^{uu} g^{ij}
\partial_u A_i \partial_u A_j + \ldots \nonumber \\
&=& \frac{r_0^2}{2 R^3 g_{SG}^2} \int du \, d^6x \, \frac{1}{u} 
\left[ A_t'^2 - f \sum_{i=1}^5 A_{x^i}'^2 \right]  
\ .
\eear
Recall from \cite{SS} the procedure for calculating
these Minkowski space Green's functions for a scalar
$\phi(u)$.
We extract the function $A(u)$ that multiplies 
$(\partial_u \phi)^2$ in the action:
\be
S = \frac{1}{2} \int du \, d^6x \, A(u) (\partial_u \phi)^2 \ .
\ee
Next, we express the bulk field $\phi$ via its value $\phi_0$
at the boundary $u=0$, $\phi(u,q) = f_q(u) \phi_0(q)$.
By definition $f_q(0) = 1$.  Moreover, we impose
an incoming-wave boundary condition on $f_q(u)$ at the
horizon $u=1$ (when $q$ is timelike).  The retarded (Minkowski space)
Green's function is then defined to be
\be
G^R(q) = \left. A(u) f_{-q}(u) \partial_u f_q(u) \right|_{u=0} \ .
\ee

Using this prescription, we find that
the retarded Green's functions for the R-current are
\bear
G_{tt}^{ab} &=&
{\mathcal C} 
\frac{ q^2 \delta^{ab}}{i\omega - D_R q^2} + \cdots
\label{Gtt} \ , \\
G_{xt}^{ab} = G_{tx}^{ab} &=&
-{\mathcal C}
\frac{ \omega q \delta^{ab}}{i\omega - D_R q^2} + \cdots
\label{Gxt} \ , \\
G_{xx}^{ab} &=&
{\mathcal C}
\frac{ \omega^2 \delta^{ab}}{i\omega - D_R q^2} + \cdots
\label{Gxx} \ , \\
G_{\alpha \alpha}^{ab} &=& -{\mathcal C} \delta^{ab}
(i\omega+2D_R q^2) + \cdots
\label{Gaa} \ . 
\eear
where
\be
{\mathcal C} = \frac{r_0^{3/2}} {R^{3/2} g_{SG}^2} \sim N^3 T^3 
\ .
\ee
and 
\be
D_R = \frac{3}{8\pi T} \ .
\label{DR5}
\ee
As expected, the $G_{\alpha \alpha}$ Green's functions have no pole
while the others do.  From the location of the pole, we can read
off the diffusion coefficient for the R-charge, $D_R$.  
We regard this value of $D_R$ as a prediction for the theory
living on a stack of M5-branes at finite temperature.  
The power of $T$ in $D_R$ is forced by dimensional analysis.
However, 
it is interesting that this value for $D_R$ is $N$ independent.

This expression for $D_R$ is subject to two kinds of
correction.  First, the supergravity approximates these M-branes
well only at large $N$.  Thus, there could be $1/N$
corrections.  The second correction is not really a correction
to $D_R$ but to the location of the pole itself.  We could
easily calculate the $A_\mu$ to higher order in $q_5$ and $\omega_5$.
In this case, we would get corrections of order ${\mathcal O}(q_5^2)$
to the location of the pole.

Although we did not solve for $g_{SG}$ exactly, we can count
powers of $N$ and $T$ in ${\mathcal C}$.  
The power of $T$ is forced by 
dimensional analysis.    
Tracing powers of $N$ is relatively easy.  The coupling
${g_{SG}}$ is essentially $\kappa_{11}$ multiplied by
lots of $N$ independent compactification factors,
and $\kappa_{11} \sim N^{-3/2}$.

\subsection{R-charge and M2-branes}

We now redo this same calculation in the nonextremal M2-brane background
(\ref{met4}).  The differential equations for $A_\mu$ take the modified
form
\bear
\omega_2 A_t' + q_2 f A_x' &=& 0 
\label{A21} \ , \\
A_t'' - \frac{1}{4f} \left(\omega_2 q_2 A_x + q_2^2 A_t \right) &=& 0
\label{A22} \ , \\
A_x'' + \frac{f'}{f} A_x' + \frac{1}{4f^2}
\left( \omega_2 q_2 A_t + \omega_2^2 A_x \right) &=& 0
\label{A23} \ , \\
A_y'' + \frac{f'}{f} A_y' + \frac{1}{4f^2}
\left( \omega_2^2 - f q_2^2 \right) A_y &=& 0 \ .
\label{A24}
\eear
where $x^0 = t$, $x^1=x$, and $x^2 = y$.  We have defined the
quantities
\be
\omega_2 \equiv \frac{3}{2\pi T} \omega \; \; \; ; \;
q_2 \equiv \frac{3}{2\pi T} q \ .
\ee
This system is very similar to the one encountered in the previous
section.  Equations (\ref{A21}) and (\ref{A22}) imply equation (\ref{A23}).
We combine equations (\ref{A21}) and (\ref{A22}) to give a single
differential equation for $A_t'$ alone:
\be
A_t''' + \frac{f'}{f} A_t'' + \frac{1}{4f^2} ( \omega_2^2 - f q_2^2 ) A_t' = 0
\label{tord}
\ .
\ee
We make the substitution $A_t' = (1-u)^{-i \omega_2 /6} F(u)$ and solve
for $F(u)$ perturbatively in $\omega_2$ and $q_2^2$:
\be
F(u) = C ( 1+ \omega_2 F_1(u) + q_2^2 G_1(u) + \ldots )
\label{Ford}
\ee
where
\bear
F_1(u) &=& \frac{i}{12} f_1(u) + \frac{i}{2\sqrt{3}} f_2(u) \ , \\
G_1(u) &=& -\frac{1}{2\sqrt{3}} f_2(u) \ .
\eear
Using (\ref{A22}), we can solve for $C$ in terms of the boundary
values of $A_t$ and $A_x$:
\be
C = \frac{\omega_2 q_2 A_x^0 + q_2^2 A_t^0}{2(i\omega_2 - \frac{1}{2}q_2^2)}
\ .
\ee
The pole in $C$ will be the same pole that appears in the R-current 
Green's functions and hence is related to the diffusion coefficient.

The perturbative expression for $A_x'$ can be obtained from
(\ref{A21}).  
We have already done the necessary work for calculating $A_y$.  Note
that (\ref{A24}) is the same differential equation as (\ref{tord}).  Thus
$A_y = (1-u)^{-i \omega_2/6} F(u)$ with $F(u)$ given by (\ref{Ford}).
The only change is that now $C = A_y^0 + {\mathcal O}(\omega_2, q_2^2)$.

To extract the Green's functions, we need to isolate the terms in
the action with two $u$ derivatives:
\be
S = \frac{r_0^2}{R^3 g_{SG}^2} \int du \, d^3x \left[
A_t'^2 - fA_x'^2 - fA_y'^2 + \ldots \right] \ .
\ee
From this expression, it is straightforward to see that
\bear
G_{tt}^{ab} &=& \frac{q^2 \delta^{ab}}{g_{SG}^2 (i\omega - D_R q^2)}
\ , \\
G_{xt}^{ab} = G_{tx}^{ab} &=& - \frac{\omega q \delta^{ab}}
{g_{SG}^2 (i\omega - D_R q^2)}
\ , \\
G_{xx} &=& \frac{\omega^2 \delta^{ab}}{g_{SG}^2 (i\omega-D_R q^2)}
\ , \\
G_{yy}^{ab} &=& -\frac{\delta^{ab}}{g_{SG}^2} 
\left( i \omega - D_R q^2 \right)
\eear
where
$1/g_{SG}^2 \sim N^{3/2}$ and the diffusion coefficient is 
\be
D_R = \frac{3}{4\pi T} \ .
\label{DR2}
\ee
This expression for $D_R$ is again $N$ independent.

\section{Stress Energy Two Point Functions}

To obtain more diffusion constants for our M-brane
theory in the hydrodynamic limit, we compute the two-point function
of the stress-energy tensor.  According to the AdS/CFT prescription,
this two-point function is related to small perturbations of the
metric in the bulk theory.  In particular, we consider
$g_{\mu\nu} = g_{\mu\nu}^{(0)} + h_{\mu\nu}$ where the 
unperturbed metric $g_{\mu\nu}^{(0)}$ 
is given by the asymptotically AdS pieces
of the full eleven dimensional metrics
(\ref{met4}) and (\ref{met7}).  
The transverse spherical
parts of the eleven dimensional metrics we leave unchanged.

For this calculation, we can ignore the spherical piece of
the metric and the four form field strength $F_4$.  They
do not enter at first order in the perturbations.  For the
asymptotically $AdS_7$ case, i.e. the five-brane case, we
consider the compactified action
\be
S = \frac{R^4 \Vol(\bS^4)}{2\kappa_{11}^2} 
\left[ \int du \, d^6x \sqrt{-g} (
{\mathcal R} - 2 \Lambda) + 2 \int d^6x \sqrt{-g^B}K \right] \ .
\ee
In the above, ${\mathcal R}$ is the Ricci scalar and $\Lambda=-15/4R^2$ is
a cosmological constant arising from the integral over $|F_4|^2$ in
the full eleven dimensional action.  The metric 
$g_{\alpha\beta}^B$ is the metric
induced on the boundary $u=0$ while $K$ is the extrinsic curvature of the 
boundary.
There is an analogous action for the asymptotically $AdS_4$ case.
\be
S = \frac{R^7 \Vol(\bS^7)}{2\kappa_{11}^2} 
\left[ \int du \, d^3x \sqrt{-g} (
{\mathcal R} - 2 \Lambda) + 2 \int d^3x \sqrt{-g^B}K \right] \ .
\ee
where 
the cosmological constant is now
$\Lambda = -12/R^2$ instead. 

The first step in computing the stress-energy two-point function is
to solve the linearized Einstein's equations
\be
{\mathcal R}^{(1)}_{\mu\nu} = \frac{2}{d-2} \Lambda h_{\mu\nu}
\ee  
where $d$ is either 4 (for asymptotically $AdS_4$) or 7 
(for asymptotically $AdS_7$).  By ${\mathcal R}_{\mu\nu}^{(1)}$, we mean
everything in the Ricci curvature 
${\mathcal R}_{\mu\nu}$ that is linear in $h_{\mu\nu}$.

The metric perturbations split up
naturally into groups as follows.  
We take a Fourier decomposition of the metric perturbation, assuming
that $h_{\mu\nu}$ depends on $t$ and $x$ as $e^{-i \omega t + i q x}$,
defining $x^0 = t$, $x^1=x$, $x^2=y$, and so on.
We choose a gauge such that $A_{\mu u}=0$.  
For the
M5-brane case, there is a rotation
group $SO(4)$ 
acting on the directions transverse to $u$, $t$, and $x$.
The perturbations split according to whether $h_{\mu\nu}$ 
is a tensor, a vector, or a scalar under these rotations.
For the asymptotically $AdS_4$ case, there is only
one direction, the $y$ direction, transverse to
the others.  However, we can still consider
the effect of sending $y \to -y$.  Under this
transformation, the metric perturbations split
into two groups according to whether $h_{\mu\nu}$
has an odd or an even number of $y$ indices.
Stretching definitions, we will call the 
metric perturbation with only one $y$ index
a vector mode.  There is one
remaining perturbation involving components
of $h_{\mu\nu}$ with both no and two $y$ indices
which we will call the scalar mode.    

The modes are useful in different ways. 
The tensor mode allows us to compute a shear
viscosity of the boundary theory using a Kubo formula.  
There is a diffusion pole in the vector mode 
which will allow us to calculate a diffusion constant.  
There is also a Kubo formula for these vector modes 
which will allow for another check of the viscosity.
The scalar mode describes sound propagation on the boundary.

We will begin by computing the diffusion constant from the 
poles in the vector modes, both for asymptotically
$AdS_4$ and $AdS_7$.  The calculation is very similar
to that of the R-charge diffusion constant.  The relation should
not be at all surprising because from an eleven
dimensional standpoint, the $A_\mu$ vector potential is
is in part a metric perturbation of the form
$h_{M \mu}$ where $M$ is a spherical direction
and $\mu$ is an asymptotically $AdS$ direction.
Next, we will analyze the tensor mode for asymptotically
$AdS_7$.   
We leave sound propagation for future work.

Sound propagation is the most involved of the three
types of metric fluctuations.  To consider
a tensor fluctuation for the M5-brane, 
$h_{yz}$ is the only
component of the 
metric fluctuations that needs to be nonzero.
For the vector fluctuations, two components
of $h_{\mu\nu}$ need to be nonzero, for example
$h_{xy}$ and $h_{ty}$.  For the scalar modes,
all of the diagonal components of $h_{\alpha\beta}$
plus $h_{xt}$ must be nonzero.  The resulting
system of differential equations is less 
tractable than for the tensor or vector modes.

\subsection{Vector Modes and a Diffusion Constant for M5-branes}

We consider a metric perturbation of the form $h_{ty} \neq 0$
and $h_{xy} \neq 0$ with all the other $h_{\mu\nu} = 0$.
Moreover, we make a Fourier decomposition such that
\bear
h_x^y &=& e^{-i\omega t + i q x} H_x(u)  \ , \\
h_t^y &=& e^{-i\omega t + i q x} H_t(u)  \ . 
\eear
The linearized Einstein's equations for $H_x$ and $H_t$ are
\bear
\omega_5 H_t' + q_5 H_x' f &=& 0 \ , 
\label{A31} \\
H_t'' - \frac{2}{u} H_t' - \frac{1}{uf} ( \omega_5 q_5 H_x + q_5^2 H_t) 
&=& 0 \ , 
\label{A32} \\
H_x'' - \frac{3-f}{uf}H_x' + \frac{1}{uf^2} (\omega_5 q_5 H_t + \omega_5^2 H_x)
&=& 0 \ . \label{A33}
\eear
This system of equations is very similar to the systems
(\ref{A1})-(\ref{A3}) and (\ref{A21})-(\ref{A23}) we solved
in the previous sections and is tractable using exactly the same
methods.  We combine the first two equations (\ref{A31}) and (\ref{A32})
to get a single equation for $H_t'$:
\be
H_t''' + \frac{2f-3}{uf} H_t'' + \frac{1}{uf^2}
(\omega_5^2 - fq_5^2 +6u^2 f) H_t' = 0 \ .
\ee
We make the ansatz $H_t' = C(1-u)^{-i\omega_5/3} F(u)$ and solve
for the regular function $F(u)$ perturbatively in $\omega_5$
and $q_5$:
\be
F(u) = u^2 + i \omega_5 \left( \frac{1}{2}(u^2-1) + \frac{1}{6}u^2 f_1(u)
+\frac{u^2}{\sqrt{3}}f_2(u)
\right) + \frac{q_5^2}{6} (1-u^2) \ . 
\ee
Taking the limit $u\rightarrow 0$, we solve for $C$ in terms of
$H_x^0$ and $H_t^0$ using (\ref{A32}):
\be
C = \frac{\omega_5 q_5 H_x^0 + q_5^2 H_t^0}{i\omega_5 - \frac{1}{3} q_5^2}
\ .
\ee
To get the two-point functions, we need to isolate the terms in the
action proportional to $H_x'^2$ and $H_t'^2$:
\be
S = \frac{2^6\pi^3}{3^7} N^3 T^6 \int du \, d^6x \frac{1}{u^2}
\left[ H_t'^2 - fH_x'^2 + \ldots \right] \ .
\ee
A subtle point can be made about the Gibbons-Hawking term here.
In order to isolate these $H_\alpha'^2$ terms in the action,
we have integrated by parts terms of the form $H_\alpha H_\alpha''$.
During this integration by parts, boundary terms of the form
$H_\alpha H_\alpha'$ appear at $u=0$.  One might think that
these boundary terms will affect the overall normalization
of the two-point functions, but they can't.  The reason
is that the Gibbons-Hawking term is precisely of a form
that cancels these particular boundary contributions.
Recall that the extrinsic curvature is defined to be
\be
K = -\nabla^\mu n_{\mu}
\ee
where $n_\mu$ is a unit vector, normal to the boundary $u=0$.
There is a piece of $K$ that looks like
\be
2 \sqrt{-g^B} K = 
\left. -\sqrt{-g} g^{uu} g^{\alpha\beta} g_{\alpha\beta,u} \right|_{u=0}
+ \ldots \ ,
\ee
which precisely cancels the boundary contribution from
terms of the form $H_\alpha H_\alpha''$.

From this quadratic piece of the action, 
we read off the retarded Green's functions:
\bear
G_{ty,ty} &=& \frac{2^5 \pi^2 N^3 T^5 q^2}{3^6 (i\omega- Dq^2)}
\ , \\
G_{ty,xy} &=& -\frac{2^5 \pi^2 N^3 T^5 \omega q}{3^6 (i\omega- Dq^2)}
\ , \\
%
G_{xy,xy} &=& \frac{2^5 \pi^2 N^3 T^5 \omega^2}{3^6 (i\omega- Dq^2)}
\ . 
\eear
where the diffusion coefficient is
\be
D = \frac{1}{4 \pi T} \ .
\label{Dgrav5}
\ee

There exists a Kubo formula (see for example \cite{KadMart}) for
these two-point functions that will let us calculate
the viscosity $\eta$.  In particular
\be
\eta = -\lim_{\omega\to 0} \left[ \lim_{q\to 0} \frac{\omega}{q^2} 
\mbox{Im}\, G_{ty,ty} \right] 
= \frac{2^5 \pi^2}{3^6} N^3 T^5 \ .
\label{Kvisc}
\ee


There is a relation between the shear viscosity (\ref{Kvisc})
and the diffusion constant (\ref{Dgrav5}):
$D = \eta / (\epsilon + P)$,
where $\epsilon$ is the energy density and $P$ is the pressure.
The relation is a consequence of the conservation condition
on the stress-energy tensor, $\partial_\alpha T^{\alpha\beta} = 0$,
along with a linearized hydrodynamic equation for the
purely spatial parts of the stress-energy tensor 
\cite{Landau}\footnote{
The bulk viscosity in these M-brane models is zero.}
\be
T^{ij} = P \delta^{ij} - \frac{\eta}{\epsilon + P}
\left( \partial_i T^{0j} + \partial_j T^{0i}
- \frac{2}{d-2} \delta^{ij} \partial_k T^{0k} \right) \ ,
\ee
where $d$ is the dimension of $AdS_d$.
Putting these two equations together, one finds a diffusion
equation for the shear modes, i.e. the fluctuations of
the momentum density $T^{0i}$ such that $\partial_i T^{0i}=0$.
The diffusion equation for $T^{ty}$ is 
$\partial_t T^{ty} = D \partial_x^2 T^{ty}$
where $D= \eta / (\epsilon + P)$.

This relation between $\eta$ and $D$ will let us check
that our calculations are internally consistent. 
From the thermodynamic relation between the entropy density
and the pressure, $S = \partial P / \partial T$, one can calculate
$P$.  Because the stress energy tensor is traceless, for the M5-branes $\epsilon = 5P$.
One finds that indeed $D = \eta / 6P = 1/4 \pi T$.

\subsection{Vector Modes and a Diffusion Constant for M2-branes}

The same story can be repeated almost verbatim for the
M2-brane case.  We take the analogous definition for
$H_x$ and $H_t$, being careful to raise the indices of
the $h_{ty}$ and $h_{xy}$ metric perturbations with 
the appropriate non-extremal M2-brane metric instead. 
The linearized Einstein's equations for $H_x$ and $H_t$ are
\bear
\omega_2 H_t' + q_2 H_x' f &=& 0 \ , 
\label{A41} \\
H_t'' - \frac{2}{u} H_t' - \frac{1}{4f} ( \omega_2 q_2 H_x + q_2^2 H_t) 
&=& 0 \ , 
\label{A42} \\
H_x'' - \frac{3-f}{uf}H_x' + \frac{1}{4f^2} 
(\omega_2 q_2 H_t + \omega_2^2 H_x)
&=& 0 \ . \label{A43}
\eear
Some small changes in the wave-vector dependent pieces of the 
differential equations are apparent, but otherwise the
system is virtually identical to that of (\ref{A31})-(\ref{A33}).
We combine the first two equations (\ref{A41}) and (\ref{A42})
to get a single equation for $H_t'$:
\be
H_t''' + \frac{f-3}{uf} H_t'' + \frac{1}{4f^2}
(\omega_2^2 - fq_2^2 ) H_t' + \frac{2}{u^2f}(3-2f)H_t'= 0 \ .
\ee
We make the ansatz $H_t' = C(1-u)^{-i\omega_2/6} F(u)$ and solve
for the regular function $F(u)$ perturbatively in $\omega_2$
and $q_2$:
\be
F(u) = u^2 + i \omega_2 \left( \frac{1}{2}(u^2-u) + \frac{1}{12}u^2 f_1(u)
-\frac{u^2 \sqrt{3}}{6} f_2(u) \right)  
+ \frac{q_2^2}{12} (u-u^2) \ . 
\ee
Taking the limit $u\rightarrow 0$, we solve for $C$ in terms of
$H_x^0$ and $H_t^0$ using (\ref{A42}):
\be
C = \frac{\omega_2 q_2 H_x^0 + q_2^2 H_t^0}
{2 \left( i\omega_2 - \frac{1}{6} q_2^2 \right)}
\ .
\ee
To get the two-point functions, we need to isolate the terms in the
action proportional to $H_x'^2$ and $H_t'^2$:
\be
S = \frac{2^{5/2}\pi^2}{3^4} N^{3/2} T^3 \int du \, d^3x \frac{1}{u^2}
\left[ H_t'^2 - fH_x'^2 + \ldots \right]
\ee
From here, we read off the retarded Green's functions:
\bear
G_{ty,ty} &=& \frac{2^{3/2} \pi N^{3/2} T^2 q^2}{3^3 (i\omega- Dq^2)}
\ , \\
G_{ty,xy} &=& -\frac{2^{3/2} \pi N^{3/2} T^2 \omega q}{3^3 (i\omega- Dq^2)}
\ , \\
G_{xy,xy} &=& \frac{2^{3/2} \pi N^{3/2} T^2 \omega^2}{3^3 (i\omega- Dq^2)}
\ . 
\eear
where the diffusion coefficient is again
\be
D = \frac{1}{4 \pi T} \ .
\label{Dgrav2}
\ee

Invoking the Kubo formula, one finds from the 
normalization of the Green's functions that the shear viscosity
is
\be
\eta = \frac{2^{3/2} \pi}{3^3} N^{3/2} T^2 \ .
\label{visc2}
\ee
We can check that the diffusion constant $D=1/4\pi T$ is consistent with
this formula for the viscosity, $D = \eta/(\epsilon + P)$.
Tracelessness of the stress energy tensor now implies
that $\epsilon = 2P$.  Using the fact that $S = \partial P / \partial T$,
one finds that indeed $D = \eta/ 3P = 1/ 4 \pi T$.

\subsection{Tensor Perturbations, M5-branes, and a Kubo Formula}

We consider a perturbation of the non-extremal M5-brane metric
of the form $h_{yz} \neq 0$ with all other $h_{\mu \nu}=0$.
We decompose such a perturbation into Fourier modes, defining
\be
h_{y}^{z} \equiv e^{-i\omega t + i q x} \phi(u) \ .
\ee
The linearized Einstein equation for $h_{yz}$ becomes
\be
\phi'' - \frac{2+u^3}{uf} \phi' + \frac{1}{uf^2} 
(\omega_5^2 - f q_5^2) \phi = 0 \ .
\ee
This differential equation can be rewritten more simply as
$\square h_y^z = 0$.  Thus this particular metric perturbation
can be thought of as a massless scalar.

We solve this differential equation in the same way we have
solved the others in this paper:  we define 
$\phi = C (1-u)^{-i\omega/3} F(u)$ and solve for $F(u)$ perturbatively
in $\omega_5$ and $q_5$.  The result is
\be
F(u) = 1 - \frac{i\omega_5}{3} f_1(u) + q_5^2 \left(
-\frac{1}{4} f_1(u) - \frac{\sqrt{3}}{6} f_2(u) \right) \ .
\ee
The constant $C$ can be re-expressed in terms of the 
boundary value of $\phi$, $C = \phi_0 / F(0)$.

To calculate the two point function, we isolate the term in
the action proportional to $\phi'^2$:
\be
S = -\frac{2^6 \pi^3 N^3 T^6}{3^7} \int du\, d^6x \frac{f}{u^2} \phi'^2 + \ldots
\ .
\ee
From this expression, we find that the retarded Green's function is
\be
G_{yz, yz} = -\frac{2^5 \pi^2 T^5 N^3}{3^6} 
\left( i\omega + \frac{3}{8\pi T} q^2 \right) \ .
\ee

According to \cite{PSS}, 
there is a Kubo formula which relates this Green's function
to the shear viscosity $\eta$.  
The formula states that
\be
\eta = -\lim_{\omega\to 0} \left[ \lim_{q\to 0}
\frac{1}{\omega} \mbox{Im}\, G_{yz, yz}
\right]
= \frac{2^5 \pi^2}{3^6} N^3 T^5 \ 
\ee
which matches (\ref{Kvisc}).\footnote{
There are still other ways of calculating the viscosity
through AdS/CFT correspondence.  For example, the authors
of \cite{visc} related the graviton absorption
cross section by non-extremal D3-branes to the viscosity
of finite temperature ${\mathcal N}=4$ $SU(N)$ super 
Yang-Mills theory.
Presumably a similar calculation could be done for
the non-extremal M2- and M5-brane supergravity solutions.}

\section{$N$-puzzles}

The viscosities and diffusion constants 
calculated here present new $N$-puzzles,
new questions about why particular $N$ dependences arise
in these M-brane theories.  From the gravitational
point of view, the $N$ dependence comes from the
gravitational coupling constant $\kappa_{11}$.
What is missing is a direct 
understanding of how a field theoretic description
of a stack of $N$ M-branes gives rise to the
same $N$ dependences.
While for D3-branes, the $N$ dependence could be 
understood perturbatively from $SU(N)$ gauge theory, 
for the M-branes, no equivalent of $SU(N)$ gauge theory
exists (as yet) as an alternate description. 
To review, the diffusion constants for the M-branes
are all $N$ independent (\ref{DR5}, \ref{DR2}, \ref{Dgrav5}, 
\ref{Dgrav2}).
The viscosities (\ref{Kvisc}, \ref{visc2}) on the other hand have the same $N$ 
dependence as the entropy (\ref{S5}, \ref{S2}) 
of the corresponding M2- or M5-brane
theory.  Note that the equation $D = \eta / (\epsilon + P)$ 
relates the $N$ dependence of the viscosity to that of the
entropy and of the diffusion constant $D$.  But there is still
at least one new unexplained $N$ dependence here.

\[
\begin{array}{l|cccc|}
{} & S & \eta & D & D_R \\ \hline
\mbox{D3-branes} & \frac{\pi^2}{2} N^2 T^3 &
\frac{\pi}{8}N^2 T^3 &
\frac{1}{4\pi T} &
\frac{1}{2 \pi T} \\
\mbox{M5-branes} & \frac{2^7 \pi^3}{3^6} N^3 T^5 &
\frac{2^5 \pi^2}{3^6} N^3 T^5 &
\frac{1}{4\pi T} &
\frac{3}{8\pi T} \\
\mbox{M2-branes} & \frac{8 \sqrt{2} \pi^2}{27} N^{3/2} T^2 &
\frac{2 \sqrt{2} \pi}{27} N^{3/2} T^2 &
\frac{1}{4\pi T} &
\frac{3}{4\pi T} \\ \hline
\end{array}
\]

The first M-brane $N$-puzzle was the entropy puzzle, the $N^3$ dependence
of the M5-brane entropy (and also the $N^{3/2}$ 
dependence of the M2-brane entropy) \cite{KlebTseyt}.  For
the D3-branes, the $N^2$ dependence was more or less clearly
the $N^2$ degrees of freedom of a $SU(N)$ gauge theory
\cite{Peet}, but
what could provide $N^3$ degrees of freedom?  

A less well known $N$-puzzle comes from the $N$ dependence of
the normalized three-point functions for M-branes.  For D3-branes,
the three-point functions scale as $N^{-1}$, as can be easily
understood from 't Hooft counting.
However, from \cite{BastZuch}, we know that the three-point functions
for M5-branes scale as $N^{-3/2}$ while for M2-branes,
as $N^{-3/4}$.  

The original motivation for writing this paper arose precisely
out of these $N$-puzzles, a feeling that these $N$ dependences
might provide a key to better understanding M-brane theories.
In particular, the more $N$ dependences we know, perhaps the better
our chances of finding some pattern and unlocking the mysteries
of M-theory.\footnote{
For another $N$-puzzle involving the conformal anomaly, see
\cite{otherpuzzles}.}

\section*{Acknowledgments}

I would like to thank Gary Horowitz, Amulya Madhav, 
Joe Polchinski,
Radu Roiban, Johannes Walcher,
and Anastasia Volovich for discussions.
I would also like to thank Igor Klebanov and especially
Dam Thanh Son for correspondence.  
Finally, I would like to thank Oliver DeWolfe for comments
on the manuscript.
This research was supported in 
part by the National Science Foundation under
Grant No. PHY99-07949.

\end{document}